\documentstyle[12pt]{article}
\def\be{\begin{equation}}
\def\ee{\end{equation}}
\def\beq{\begin{eqnarray}}
\def\eeq{\end{eqnarray}}
\begin{document}
\hoffset 0.5cm
\voffset -0.4cm
\evensidemargin 0.0in
\oddsidemargin 0.0in
\topmargin -0.0in
\textwidth 6.1in
\textheight 8.7in

\begin{titlepage}

\begin{flushright}
BRX-TH-394\\
hep-th/9604156\\
April 1996
\end{flushright}

\vskip 0.2truecm

\begin{center}
{\large {\bf Comparing D-branes to Black-branes$^*$}}
\end{center}

\vskip 0.4cm

\begin{center}
{Gilad Lifschytz}
\vskip 0.2cm
{\it Department of Physics,
     Brandeis University,\\
     Waltham, MA 02254, USA.\\ 
    e-mail: lifschytz@binah.cc.brandeis.edu }

\end{center}

\vskip 0.8cm

\noindent {\bf Abstract} We calculate the potential between two different
stationary D-branes and the velocity dependent potential between 
two different moving D-branes. 
We identify configurations with some unbroken supersymmetry, 
using a zero force condition.
The potentials are compared with an eleven dimensional 
calculation of the scattering of a zero 
black-brane from the $0,2,4$ and $6$
black-brane of type IIA supergravity. The agreement of these calculations
provide further evidence for the D-brane description of black-branes,
and for the eleven dimensional origin of type IIA string theory.

\noindent                
\vskip 0.8 cm

\begin{flushleft}

$^*$ {\small This work was supported 
by the National Science Foundation grant PHY-9315811.}

\end{flushleft}

\end{titlepage}

\section{Introduction}
It has been recently realized that 
string dualities can be understood in a simple way by assuming the 
existence of higher dimensional theories \cite{wit,sch,vaf}. 
In particular an eleven dimensional 
``M''-theory is conjectured to give, when compactified on different 
manifolds, various string theories in ten dimensions 
\cite{wit,horwit,polwit}. The relationship of the M-theory
to type IIA string theory is particularly simple. The M-theory on 
$R^{10} \times S^1$ is conjectured
to be equivalent to type IIA string theory with a particular string coupling
($g_s$), which is determined by the radius of $S^1$ 
(which we denote by $R_{11}$), $g_s \sim R_{11}^{3/2}$.
The low energy limit of the M-theory should be 11D supergravity
\footnote{
It has been known for a long time that
the type IIA supergravity can be obtained by a dimensional reduction of
eleven dimensional supergravity on $R^{10} \times S^1$.}.

The Kaluza-Klein mode of the 11 dimensional theory is 
a non perturbative state in type IIA string theory. Other non-perturbative
relationships between string theories require extended 
objects with particular properties \cite{wit,hultow,sen}. 
These object have been conjectured to be the D-branes \cite{pol}
(for a recent review see \cite{cjp}).
The type IIA and type IIB supergravity theories have black-brane solutions.
These are extended objects with horizons and singularities. Some of these
solutions carry RR charges and in the extremal limit preserve half of the
space-time supersymmetries. These were thought to be the extended objects
needed for string dualities before the D-brane idea came. 
Now they are believed
to be the long range space-time description of the D-branes. 
Thus, the description of the D-branes in terms of boundary conditions
of open strings might give us a powerful tool to analyze the behavior
of the black-branes.
In fact, the D-brane description of black-branes 
(if correct) may be used to
solve some long standing problems with black holes. It has been used to
calculate the entropy
of black holes in various dimensions \cite{bh}, and to try and resolve the
black hole paradox. Thus it
has become important to test the equivalence between the D-branes and the
black-branes.                                   
It has been shown \cite{ghkm,gamy}, that string scattering
off D-branes agrees with the results (at large distances) 
of scattering particles off the black-branes, showing that they 
carry the same charges.

The purpose of this letter is to give further evidence that the long range
space-time description of the D-branes are the black-branes, and to 
show that
the long range interaction between the D-branes follows from eleven
dimensional physics (see also \cite{asy}).
In this letter we  
compute the potential between two different stationary branes,
and between two different moving branes using the result from
\cite{pol,bac}. We then compute the potential between the black-branes of
type IIA supergravity and a moving zero black-brane. 
This calculation is done
using an eleven dimensional interpretation of these solutions \cite{town}.

The D-brane calculation is valid for small string coupling which is
equivalent from the point of view of 11D to a small radius of the eleventh
direction. 
In general an eleven dimensional computation might be expected to
be valid only at strong coupling. However we are going to use the 
eleven dimensional
interpretation only to determine the form of the interaction 
between the ten
dimensional fields of the black-branes. When the M-theory is compactified
on $R^{10} \times S^1$ with a small $R_{11}$ we know that the only massless
particles are the ones from the reduction of the 11D supergravity down to
ten dimensions, so no new long range physics is involved.
Thus if we interpret the black-branes as some solutions of the low energy 
supergravity in 11D, and derive the 
interaction between them from 11D physics,
as long as we are looking at the long range interaction and at processes
that do not excite 11D modes, they can be compared with a D-brane 
calculation.
Comparing both calculations we find that they agree at large
distances. The agreement of the two
calculations gives further evidence to the consistency 
of the eleven dimensional
picture.

\section{Potential between D-branes}
In this section we compute the static potential between two 
different D-branes \cite{dlp,gr}. 
The vanishing of the potential between two different 
D-branes is a signature of the
existence of some unbroken supersymmetries in the D-branes configuration.

A D-brane of dimension $p$ is defined by the boundary condition 
of open strings,
Neumann boundary conditions on $p+1$ coordinates, 
\be
n^a \partial_a X^{\mu} =  0 \ \ \mu =0, \ldots ,p,
\ee
and Dirichlet boundary conditions on $9-p$ coordinates,
\be
X^{\mu} =  0 \ \ \mu=p+1, \ldots ,9.
\ee
We define the first D-brane to be 
of dimension $p$ and the second of dimension $l$, where we take $p \geq l$.
The branes coordinate are either parallel or orthogonal and we label the
number of orthogonal coordinates as $a$ 
(of course $l \geq a $, $p+a \leq 9$ ).

We follow \cite{pol} and compute the effective potential of a stretched open  
superstring between the two branes. 
Now the string coordinate can have three different types of
boundary conditions
on its two ends (the boundary condition are enforced at $\sigma=0$
and at $\sigma=\pi$): 
NN, DD and ND, where N (D) stands for Neumann (Dirichlet) 
boundary conditions.
The branes are separated by a distance $b$, which
is taken to be in one of the DD directions.
The number($\sharp$) of string coordinates of each type is,
\begin{enumerate}
\item $\sharp DD = 8-p-a$
\item $\sharp NN = l-a$
\item $\sharp ND = p-l+2a$
\end{enumerate}
Where we have taken into account that the ghost contribution will cancel
two of the coordinates (NN or DD). 

In order to calculate the potential between the D-branes we
calculate the one loop vacuum amplitude of open 
superstring in the presence of
the D-branes \cite{pol}. The one loop vacuum amplitude takes the form,
\begin{equation}
A=C\int_{0}^{\infty} \frac{d^{(\sharp NN+1)} k}{(2\pi)^{(\sharp NN+1)}} 
\sum_{i} \int_{0}^{\infty} \frac{dt}{t} e^{-2\pi\alpha' t(k^2 +M_{i}^{2})}
\end{equation}
where $\frac{1}{2\pi \alpha'}$ is the string tension, 
the sum runs over all the string states, 
\begin{equation}
M_{i}^{2}=\frac{b^2}{4\pi^2 \alpha'^2}+ \frac{1}{\alpha'}\sum (oscillators)
\end{equation}
and $C$ is the space time volume of an $(l-a)$ brane. Integrating over $k$
and doing the oscillator sum gives
\begin{equation}
A=C\int \frac{dt}{t} e^{-(\frac{b^2 t}{2\pi \alpha'})} 
(8\pi^2 \alpha' t)^{-(\sharp NN +1)/2} B \times J.
\end{equation}
$B$ and $J$ are the contribution from the bosonic and fermionic oscillators
respectively.

For the bosonic oscillators the NN and DD boundary
condition gives as usual integer modes, while the ND boundary
condition gives the bosonic modes a spectrum of half integers.
For the fermionic oscillators in the NS (Neveu-Schwarz) 
sector the NN and DD are as usual half integer modes, and
ND boundary conditions gives integer modes. 
In the R (Ramond) sector NN and DD are integer modes while ND 
gives half integer modes. The degeneracy of the ground states becomes
$2^{\sharp(NN+DD)/2}$ in the R sector, 
and $2^{\sharp ND/2}$ in the NS sector.
If $\sharp ND \neq 0$ there are fermionic 
zero modes in the NS$(-1)^F$ sector,
and thus this sector does not contribute
\footnote{The case $\sharp ND=8$ is special and will be treated later}
(like the R$(-1)^F$).
In terms of $q=e^{-\pi t}$, we define
\begin{eqnarray}
f_{1}(q) & = & q^{1/12} \prod_{n=1} (1-q^{2n}). \\
f_{2}(q) & = & \sqrt{2} q^{1/12} \prod_{n=1} (1+q^{2n}). \\
f_{3}(q) & = & q^{-1/24} \prod_{n=1} (1+q^{2n-1}). \\
f_{4}(q) & = & q^{-1/24} \prod_{n=1} (1-q^{2n-1}).
\end{eqnarray}
Then the bosonic and fermionic contributions are,
\be
B=f_{1}^{-\sharp (NN+DD)}(q)f_{4}^{-(\sharp ND)}(q),
\label{stbos}
\ee
\be
J=\frac{1}{2}\{-f_{2}^{\sharp(NN+DD)}(q)f_{3}^{\sharp ND}(q) + 
f_{3}^{\sharp(NN+DD)}(q)f_{2}^{\sharp ND}(q) \}.
\label{stfer}
\ee

In order to calculate the long range potential one needs 
the expansion of the $f_{i}(q)$ functions as $t \rightarrow 0$.
\beq
f_{1}(q) & \rightarrow & \frac{1}{\sqrt{t}} e^{-\pi/(12t)}. \\
f_{2}(q) & \rightarrow & e^{\pi/(24t)}(1-e^{-\pi/t}). \\
f_{3}(q) & \rightarrow & e^{\pi/(24t)}(1+e^{-\pi/t}). \\
f_{4}(q) & \rightarrow & \sqrt{2} e^{-\pi/(12t)}. 
\eeq
Inserting those expressions into equation (\ref{stfer},\ref{stbos}) one gets
\beq
B & = & 2^{-\sharp ND/2}t^{\sharp (DD+NN)/2}\exp{(\frac{8\pi}{12t})}. \\
J & = & \sharp(DD+NN-ND)\exp{-(\frac{8\pi}{12t})}. 
\eeq
Then the one loop vacuum amplitude, in the limit of large $b$, is
\be
A=C[2-(p-l)/2+a] T_p T_l G_{9-p-a}(b^2),
\label{fstat}
\end{equation}
where $T_i=\sqrt{\pi}(4\pi^2 \alpha')^{(3-i)/2}$ is the tension 
of an  i-brane \cite{pol}, and 
$$G_{9-p-a}=\frac{1}{4} \pi^{(p-9+a)/2} 
\Gamma ((7-p-a)/2) (b^2)^{(p-7+a)/2}$$
is the scalar Green function in $(9-p-a)$ dimensions.

From  equation (\ref{fstat}) the potential 
can be read off ($b^2=R^2$) to be
\be
V(R) \sim -[2-(p-l)/2+a] R^{-(p+a-7)}.
\ee
The absence of the sectors NS$(-1)^F$ and R$(-1)^F$ reflects the fact that 
different D-branes do not interact through their RR gauge fields.
As one can see from equations (\ref{stbos},\ref{stfer})
the interaction between two D-branes is governed by the number of
$ND$ type coordinate, which is a T-duality invariant quantity.

From equation (\ref{stfer}) one sees that the static potential vanishes
for $\sharp ND=4$. In these cases the two branes do not exert force
on each other and thus 
these configuration preserves some supersymmetry. 
The only possibilities (see also \cite{cjp,grgu}) are the
sequences ($p=4+l$, $a=0$), ($p=l+2$ ,  $a=1$) and ($p=l$ , $a=2$). 
The space time description of some of these configuration
have been constructed recently \cite{tsy,paptow,gkt}.

The case $\sharp ND=8$ is slightly different. Here all the modes in the R
sector are half integer and all the modes in the NS sector are integers.
In this case there are no fermionic zero modes in the  R$(-1)^F$ sector
 and this sector contributes. One finds, 
\be
A \sim \int \frac{dt}{t} e^{-\frac{b^2 t}{2\pi \alpha}} 
(8\pi^2 \alpha t)^{(\sharp DD-1)/2} B \times J.
\ee
\be
B\times J=\frac{1}{2}f_{4}^{-8}\{-f_{3}^{8}(q) 
+ f_{2}^{8}(q)+ f_{4}^{8}(q)\}
\label{specase}
\ee
In the above expression, if $\sharp DD +1 =0 $ we should set $b^2 =0$.

Equation (\ref{specase})  vanishes due to the `abstruse identity',
This means that a $p$ and an $l$ brane
in a configuration with $p-l+2a=8$ (with $p+a \leq 9$) 
do not exert force on each other. 
Thus this configuration preserves some supersymmetry. 
Some examples are the string and the nine-brane, the zero-brane and the
eight-brane, the seven-brane and the D-instanton, 
two five-branes intersecting on a common string, two
totally orthogonal four-branes, etc. 
(again in agreement with \cite{cjp,grgu}). If one of the above branes is
an anti-brane then the sign of the factor $f_{4}^{8}(q)$ in equation 
(\ref{specase}) changes, and
there is no cancellation.
The space time description of these
configurations can be obtained using the rules in \cite{tsy}. 
In fact
the metric for the two five-branes intersecting on a common string is
presented in \cite{gkt} and all the rest can be obtained using T-duality
transformations of \cite{bho,brgpt}.

One can also identify when a tachyonic instability can arise \cite{bansus}.
This happens when the ground state energy in the NS sector is negative
which is when $\sharp(ND-NN-DD) <0$.

\section{Moving D-branes}
In this section we follow \cite{bac}. We will only consider the case $a=0$.
We investigate the potential between
two relatively moving different branes of dimensions
$p,l$ respectively,  
where the $l$ brane moves with velocity $v$ with respect
to the $p$ brane in one of the DD directions (taken to be $X_d$), 
i.e $X_{d}=vX_{0}$. 
As before the branes are separated in another DD 
coordinate by a distance $b$.
As was shown in \cite{bac} the correction to the vacuum amplitude due to the
moving of the branes
comes from the ratio of the contribution of the ($X_0$,$X_d$) coordinates,
and those of the ghost. 
Similarly for the ratio of the fermionic coordinates 
and the super-ghosts. If the branes are stationary the ratio is $1$. 
The ratio is different, when the branes 
are moving with respect to each other,
because the boundary condition for $X_0$  
($X_d$) are not NN (DD) any more. Rather they are 
\beq
X_d -v X_0 & = & \partial_{\sigma}(v X_d - X_0)=0 \ \ (at\ \ \sigma=0). \\
X_d & = & \partial_{\sigma} X_0 =0 \ \ (at\ \ \sigma=\pi).
\eeq  
Similarly the boundary condition on the fermionic 
coordinates is also changed
\cite{bacpor}.
The one loop vacuum amplitude, $A$, takes the form
\begin{equation}
A=2\frac{C'}{4\pi}\int_{0}^{\infty}\frac{dt}{t}(8\pi^2 \alpha' t)^{-l/2}
 e^{-(b^2t/2\pi \alpha')} B \times J
\label{movd}
\end{equation}
where $C'$ is the space volume of the $l$ brane.
The bosonic and fermionic parts are 
($\Theta_i$ are the usual Jacobi functions,
and prime denotes the derivative with respect to the first argument.),
\begin{equation}
B=f_{1}^{-\sharp(NN+DD)}(q)f_{4}^{-\sharp ND}(q) 
\frac{\Theta_{1} '(0,it)}{\Theta_{1}(\epsilon t,it)}.
\label{movbos}
\end{equation}
\begin{equation}
J=\frac{1}{2}\{-f_{2}^{\sharp(NN+DD)}(q)f_{3}^{\sharp DN}(q) 
\frac{\Theta_{2} (\epsilon t,it)}{\Theta_{2}(0,it)} + 
f_{3}^{\sharp(NN+DD)}(q)f_{2}^{\sharp DN}(q) 
\frac{\Theta_{3} (\epsilon t,it)}{\Theta_{3}(0,it)}\}
\label{movfer}
\end{equation}
and $\tanh{\pi \epsilon}=v$.
 
We are interested in the velocity dependent 
long range potential between the
branes, so we evaluate equation (\ref{movbos},\ref{movfer})
 in the limit of small velocities and small $t$.
In order to find the correction due to the velocity one uses the differential
equation solved by the Jacobi functions, 
\be
\frac{\partial ^{2}\Theta_i (\nu,\lambda)}{\partial ^{2} \nu} = 
4i\pi\frac{\partial \Theta_i(\nu, \lambda)}{\partial \lambda} 
\label{thfu}
\ee
and the known asymptotic behavior of the theta functions at $\nu = 0$.
In the limit of small velocities and small $t$  one finds,
\begin{equation}
B=2^{-(p-l)/2} t^{\frac{8-p+l}{2}} e^{8\pi/12t} 
\frac{1}{\epsilon t (1+(\pi \epsilon)^{2}/6)},
\label{corbos}
\end{equation}

\begin{equation}
J=e^{-8\pi/12t}\{(8-2p+2l)+4(\pi \epsilon)^{2} \}.
\label{corfer}
\end{equation}
The bosonic corrections are not corrections to the potential, 
but rather to the kinematics. 
collecting all the expressions together gives
\begin{equation}
A=C T_{p} T_{l}[2-(p-l)/2 +(\pi \epsilon)^{2}]
\int_{-\infty}^{\infty} d\tau G_{9-p}(b^2 +\tau^2 \sinh^{2}(\pi \epsilon))
\label{movpotf}
\end{equation}                                                        
In the above expression we have set 
$\pi\epsilon(1+\frac{(\pi\epsilon)^2}{6}) \approx \sinh(\pi \epsilon)$.
One can see that we should interpret
$\tau$ as the time in the frame of the moving brane.
From this expression one can read off the velocity correction
to the potential between an $l$-brane and a $p$-brane
($\pi \epsilon \approx v$, $R^2=b^2 +\tau^2 \sinh^{2}(\pi \epsilon)$)
\be
V(R) \sim -R^{p-7} [2-(p-l)/2+v^2].
\label{ddp}
\ee
Equation (\ref{ddp}) only depends on $v^2$. In fact equation (\ref{movfer})
is even in $\epsilon$, so one might wonder where are the linear velocity
correction to the force, expected in the case of $p+l=6$, 
due to the Lorentz force .
However the way one finds the potential is through energy considerations. 
It can't be expected to reproduce a Lorentz type force.

When $p=l$ one has \cite{bac}
\be
B=f_{1}^{-8} \frac{\Theta_{1} '(0,it)}{\Theta_{1}(\epsilon t,it)},
\label{smovbos}
\ee
\be
J=\frac{1}{2}\{-f_{2}^{8} 
\frac{\Theta_{2} (\epsilon t,it)}{\Theta_{2}(0,it)} + 
f_{3}^{8} 
\frac{\Theta_{3} (\epsilon t,it)}{\Theta_{3}(0,it)}\} \pm
f_{4}^{8}
\frac{\Theta_{4} (\epsilon t,it)}{\Theta_{4}(0,it)}\}.
\label{smovfer}
\ee
Here $+$ or $-$ are for the case of brane- anti brane and brane-brane 
scattering respectively.
As before in the long range small velocity limit, one gets
for the brane anti-brane scattering :
\be
B = t^{4} e^{8\pi/12t} 
\frac{1}{\epsilon t (1+(\pi \epsilon)^{2}/6)},
\label{sbbos}
\ee
\be
J=8e^{-8\pi/12t}(2+(\pi \epsilon)^2).
\label{babfer}
\ee
For the brane-brane case
\be
J=(\pi \epsilon)^4 e^{-8\pi/12t},
\label{bbfer}
\ee
and $B$ as in equation (\ref{sbbos}).
So the velocity dependent potentials between a zero-brane and an 
anti-zero-brane, and between two zero-branes are, 
\beq
V(R) & \sim & -R^{p-7} (2+v^2). \label{dzaz}\\
V(R) & \sim & -R^{p-7} v^{4}/8. \label{dzz} 
\eeq

\section{Scattering of black-branes}
In this section we will calculate the potential between
a moving zero black-brane and the
$0,2,4,6$ black-branes of type IIA supergravity.
The zero black-brane has
a very simple interpretation in eleven dimensions, as a KK mode. 
It is then easy to understand its
coupling to the other branes from an eleven dimensional point of view.

One starts with the extremal zero black-brane solution 
(in the string metric) 
of \cite{horstr}. 
The space-time coordinates will be labeled $1, \ldots ,11$, 
time being the first coordinate. 
The relationship with the 11D metric is \cite{wit},
\be
ds_{11}^{2}=e^{-2\phi /3}ds_{10}^{2}(string)
+e^{4\phi /3}(dx_{11}^{2}-A_{\mu}dx^{\mu})^{2}
\label{elten}
\ee
and we ignore all of the gauge potentials other that the one-form, as
they will not enter in our discussion.
In equation (\ref{elten}) $\phi$ is the dilaton and $A_{\mu}$ is the 
one-form gauge potential in the RR sector. 
Then by a change of coordinates ($R^2=\sum_{j} x_{j}^{2}$)
$R^{7-p}=r^{7-p}-r_{+}^{7-p}$, one arrives at
the following metric for the zero black-brane ($j=2, \ldots ,10$)
\be
ds_{11}^{2}=-(1-\frac{r_{+}^{7}}{R^{7}})dt^2+2\frac{r_{+}^{7}}
{R^{7}}dtdx_{11}
+(1+\frac{r_{+}^{7}}{R^{7}})dx_{11}^{2}+dx_j dx^j.
\label{zebr}
\ee

For the two black-brane (where $j=4, \ldots ,10$ and $i=2,3$),
\be
ds_{11}^{2}=(1+\frac{r_{+}^{5}}{R^{5}})^{1/3} \left[
(1+\frac{r_{+}^{5}}{R^{5}})^{-1}
(-dt^2+dy_{i}dy^{i})+dx_j dx^j +dx_{11}^2 \right].
\ee
For the four black-brane (where $j=6, \ldots ,10$ and $i=2, \ldots ,5$),
\be
ds_{11}^{2}=(1+\frac{r_{+}^{3}}{R^{3}})^{2/3} \left[
(1+\frac{r_{+}^{3}}{R^{3}})^{-1}
(-dt^2+dy_{i}dy^{i}+dx_{11}^2)+dx_j dx^j \right].
\ee
For the six black-brane
\beq              
ds_{11}^{2} & = & (1+\frac{r_{+}}{R}) \left[(1+\frac{r_{+}}{R})^{-1}
                  (-dt^2+dy_{i}dy^{i})+ \right.\nonumber \\
            &   & \left. (1+\frac{r_{+}}{R})^{-2}(dx_{11}+r_+ (1-\cos \theta)
                  d\phi)^{2}+dR^{2}+R^2d^2 \Omega_2 \right].
\eeq

Let us focus on the zero brane. Equation (\ref{zebr})
is the metric of a plane-fronted gravitational wave
moving in the $x_{11}$ direction. In fact this metric can be obtained
by starting with the Schwarzschild metric in 10D and boosting it in the
eleventh direction to the speed of light while keeping the combination
$r^{7}_{+} \sinh^{2} \alpha $ fixed (here $r_+, \alpha $ are the Schwarzschild
radius and the boost parameter respectively) \cite{hull,gibb}.
This shows that the zero black-brane metric is not a solution of
11D gravity, but rather requires a source which might be thought of as
coming from the elementary membrane of the M-theory. The fact that 
zero D-branes do not exert forces on each other, is the well known fact
that two parallel moving null particles 
do not interact.
It is represented in
the zero black-brane metric by the fact that $dx_{11}=-dt$ is a geodesic of
the metric (\ref{zebr}).
Of course
$dx_{11}=dt$ is not a geodesic of the metric (\ref{zebr}), 
which corresponds
to an interaction between a zero-brane and an anti zero-brane.

At large distances the zero black-brane does not affect the 
other black-brane's
metric, as its metric coefficient falls much faster than the 
other black-branes.
From an eleven dimensional point of view the zero 
black-brane metric is just
the metric generated by a massless scalar particle moving in the eleventh
direction.
The scattering of the zero black-brane from the other black-branes
should then be dictated by 11D diffeomorphism invariance.
To calculate the semiclassical scattering of the zero black-brane 
from the other black-branes all we need is to calculate the scattering
of the null geodesic off the other black-branes metrics.
As was explained in the introduction, 
we do expect to be able to compare this
calculation,
to a D-brane calculation, as long as we look at the long range interactions.

The potential can be basically read off the black branes metric, and
one gets for the $2,4$ and $6$  black-brane
\beq
V_{(0-2)}(R) & = & -a_2 R^{-5}(1+v^2). \label{zt} \\
V_{(0-4)}(R) & = & -a_4 R^{-3}v^2. \\
V_{(0-6)}(R) & = & -a_6 R^{-1}(v^2-1).
\eeq
In the case of the six-black-brane we have constrained the geodesic to be
a line of constant angle $\phi$,
in order to suppress the Lorentz force.

Although the above reasoning does not necessarily apply when we scatter two
zero black-brane we will compute the potential for these cases also.
For the anti zero black-brane
\be
V(R)=-a_0 2R^{-7}(2+v^{2})
\ee
and for the zero black-brane
\be
V(R)= -a_0 R^{-7}v^{4}/4.
\label{zz}
\ee
The absence of $v^{2}$ term in the potential between two identical parallel
extremal black-branes was proved in \cite{km}.

Comparing the potential derived from the 11D interpretation
of the black-branes (equations (\ref{zt}-\ref{zz})) 
to the potential computed
in the D-brane approach (equations (\ref{ddp},\ref{dzaz},\ref{dzz})), we
see that there is perfect agreement between them.

\section{Discussion}
We have shown that the computation of the potentials between two D-branes
agrees with a computation from a space-time, black-brane approach. 
The interaction
between the black-branes was taken to be the 
interaction coming from their 11D
interpretation (although only a subclass of black-branes was considered). 
We believe this is further evidence for the existence of an underlying
eleven dimensional theory.

The computation of geodesics on a given background is equivalent to the 
eikonal approximation,
which might be expected to work for the 
scattering of the zero-brane from the
anti zero-brane \cite{eik}, but it is unclear why it reproduces the
result for the zero-brane, zero-brane scattering.
\begin{center}
{\bf Acknowledgments}
\end{center}
I would like to thank S. Deser, G. Esposito-Farese and S. Mathur for
many interesting discussions.

\end{document}